\documentclass[10pt,conference]{IEEEtran}

\ifCLASSINFOpdf
\else
\fi

\usepackage[letterpaper, left=0.625in, right=0.625in, top=0.75in, bottom=0.95in]{geometry}

\ifCLASSOPTIONcompsoc
  \usepackage[caption=false,font=normalsize,labelfont=sf,textfont=sf]{subfig}
\else
  \usepackage[caption=false,font=footnotesize]{subfig}
\fi
\usepackage{xcolor}
\usepackage{lipsum}
\usepackage[T1]{fontenc}% optional T1 font encoding
\usepackage{amsmath,amsfonts}
\usepackage{tikz-cd}
\usepackage{amssymb}
\usepackage{tikz}
\usetikzlibrary{positioning, shapes, arrows.meta}
\usepackage{algorithm}
\usepackage{algpseudocode}
\usepackage{array}
\usepackage{makecell}
\usepackage{amsthm}
\usepackage{adjustbox}
\usepackage{stackengine}
\usepackage{tabularx}
\usepackage{amssymb}

\usepackage{amsmath} 
\usepackage{xcolor}
\usepackage{textcomp}
\usepackage{stfloats}
\usepackage{amsmath}
\usepackage{url}
\usepackage{verbatim}
\usepackage{graphicx,booktabs,multirow,bm}
\usepackage{subfig}
\usepackage{xcolor}
\usepackage{scalerel}
\usepackage{cite}
\usepackage{colortbl}
\newcommand{\bb}[1]{\textbf{#1}}
\newcommand{\uu}[1]{\underline{#1}}
\newtheorem{theorem}{Theorem}

\newcommand{\x}{\boldsymbol{x}}

\newcommand{\z}{\boldsymbol{z}}

\newcommand{\y}{\boldsymbol{y}}

\newcommand{\h}{\boldsymbol{h}}

\newcommand{\n}{\boldsymbol{n}}
\newcommand{\hz}{\hat{\boldsymbol{z}}}

\newcommand{\hZ}{\hat{Z}}

\IEEEoverridecommandlockouts
\begin{document}

% \title{Distribution Shift in Task-Oriented Communication:\\ An Information Bottleneck Solution}
% \title{Robust Distributed Co-Inference in Task-Oriented Communication with Information Bottleneck}
% \title{Accelerating Remote Training in Task-Oriented Communication: A Task-Agnostic Mutual Information Maximization Approach}
\title{Remote Training in Task-Oriented Communication: Supervised or Self-Supervised with Fine-Tuning?}

%% Multi-view
%% Robust
%% Real-time
%% Unsupervised
%% Task-oriented communication

% \title{Tackling Distribution Shifts in Task-Oriented Communication}
\author{\IEEEauthorblockN{Hongru~Li, Hang~Zhao, Hengtao~He, Shenghui~Song,\\ Jun Zhang,~\textit{Fellow,~IEEE,} and {Khaled B.~Letaief,~\textit{Fellow,~IEEE}}\\\IEEEauthorblockA{Dept. of Electronic and Computer Engineering,
The Hong Kong University of Science and Technology, Hong Kong\\
Email: \{hlidm, hzhaobi\}@connect.ust.hk, \{eehthe, eeshsong, eejzhang, eekhaled\}@ust.hk}}\\ \vspace*{-12mm}\thanks{This work was supported in part by the Hong Kong Research Grants Council under the Areas of Excellence scheme grant AoE/E-601/22-R, in part by NSFC/RGC Collaborative Research Scheme under grant CRS\_HKUST603/22, and in part by NSFC/RGC Joint Research Scheme under grant N\_HKUST656/22.}}

% \thanks{The authors are with the Department of Electronic and Computer Engineering, The Hong Kong University of Science and Technology (HKUST), Hong Kong (e-mail: hlidm@connect.ust.hk, jiawei.shao@connect.ust.hk, eehthe@ust.hk, eeshsong@ust.hk, eejzhang@ust.hk, eekhaled@ust.hk). (The corresponding author is Hengtao He.)}}



\maketitle

\pagestyle{plain}  % no page number for the second and the later pages
\thispagestyle{empty}
\begin{abstract}
Task-oriented communication focuses on extracting and transmitting only the information relevant to specific tasks, effectively minimizing communication overhead. Most existing methods prioritize reducing this overhead during inference, often assuming feasible local training or minimal training communication resources.
However, in real-world wireless systems with dynamic connection topologies, training models locally for each new connection is impractical, and task-specific information is often unavailable before establishing connections. Therefore, minimizing training overhead and enabling label-free, task-agnostic pre-training before the connection establishment are essential for effective task-oriented communication. In this paper, we tackle these challenges by employing a mutual information maximization approach grounded in self-supervised learning and information-theoretic analysis. We propose an efficient strategy that pre-trains the transmitter in a task-agnostic and label-free manner, followed by joint fine-tuning of both the transmitter and receiver in a task-specific, label-aware manner. Simulation results show that our proposed method reduces training communication overhead to about half that of full-supervised methods using the SGD optimizer, demonstrating significant improvements in training efficiency.
\end{abstract}

\vspace{0.2cm}
\section{introduction}
\label{sec:introduction}

Driven by groundbreaking advancements in artificial intelligence (AI) technologies, next-generation wireless networks are anticipated to deliver a wide range of advanced services, including immersive augmented reality (AR) and virtual reality (VR) experiences, intelligent transportation systems, and support for the Internet of Things (IoT)\cite{letaief2019roadmap}. However, this AI-driven evolution brings unprecedented challenges to existing communication systems\cite{shi2020communication}, particularly in managing massive data transmission and reducing the communication overhead to support these emerging applications.

% meeting ultra-low latency requirements, underscoring an urgent need for efficient communication strategies to support these emerging applications.

A promising solution to address the challenge is task-oriented communication~\cite{vfe,li2024tackling, xu2023edge}, also known as semantic communication~\cite{xie2021deep,xin2024semantic}, which can significantly reduce the communication overhead compared with traditional data-oriented communication systems. The transceivers in previous data-oriented communication aims to deliver the raw information with high fidelity, which is not always necessary for the downstream tasks and has high communication overhead. 
In contrast, the transmitter in task-oriented communication focuses on extracting and transmitting only the task-relevant information from the input data and ignoring the task-irrelevant information, while the receiver focuses on leveraging the received information to accomplish the downstream tasks.

% which significantly reduces the communication overhead.

% Existing works have shown great success of task-oriented communication in efficient information transmission especially for reducing the communication overhead to accomplish the downstream tasks.
Existing works have demonstrated the success of task-oriented communication in reducing communication overhead. The authors of~\cite{bottlenet} proposed a model splitting method to reduce communication overhead by increasing on-device processing within a deep learning-based joint source-channel coding framework~\cite{jssc_deniz}. To further reduce the transmitted symbol rate, this framework was combined with the information bottleneck (IB) principle in~\cite{vfe}, where a sparse-inducing latent prior with a feature pruning scheme based on signal-to-noise ratio (SNR) was proposed to selectively omit certain feature dimensions during transmission.
A more refined mechanism for explicitly reducing bit rate was introduced in~\cite{yang2024swinjscc}, where an additional rate control network was trained to adapt to varying channel conditions and transmission rates. It can achieve a flexible tradeoff between the performance and communication overhead. 
% Instead of feature pruning or rate control, the authors in~\cite{jiang2022wireless} proposed an incremental redundancy hybrid automatic repeat request scheme to improve the inference time reliability of task-oriented communication systems. An error detector was trained to determine which part of task-relevant information need to be retransmitted to ensure the system reliability.
Instead of feature pruning or rate control, an incremental redundancy hybrid automatic repeat request scheme was proposed in~\cite{jiang2022wireless} for efficient information retransmission. In particular, an error detector was trained to determine which part of task-relevant information needed to be retransmitted and thus the reliability of task-oriented communication systems was improved.

% An error detector was trained to determine which part of task-relevant information needed to be retransmitted to ensure system reliability.

However, existing works only consider the inference communication overhead of task-oriented communication, overlooking the communication overhead durinig training. In particular, local training is assumed to jointly train the transmitter and receiver in a centralized manner before deployed separately to the transceivers. However, this assumpation is not feasible in practical wireless communication systems as the connection topology among transceivers is dynamic, hindering locally traininig the model for each new connection~\cite{zhang2022toward}. Thus, remote training is more practical in real-world scenarios, which requires the exchange of forward intermediate features and backward gradients between the transmitter and receiver over wireless channel. Such remote training incurs significant communication overhead, and thus reducing the communication overhead during training is even more critical than that during inference.

To reduce communication overhead during training, we propose a mutual information maximization approach based on self-supervised learning and information-theoretic analysis. Self-supervised learning, which leverages intrinsic patterns in data to create useful representations without relying on corresponding label, enables us to pre-train models in a task-agnostic and label-free way\footnote{We note that a concurrent work~\cite{gu2024self} also employs self-supervised learning techniques, but with the goal of addressing the challenge of training with limited labeled data, which differs from the main focus of our proposed method.}. Specifically, we develop an efficient strategy that first pre-trains the transmitter using self-supervised learning, allowing it to learn from raw data alone. This pre-training is followed by joint fine-tuning of the transmitter and receiver in a task-specific manner using supervised learning. Simulation results demonstrate that the proposed approach significantly reduces communication overhead during training while achieving comparable or even superior performance to that of state-of-the-art supervised methods. 
% Simulation results demonstrate that the proposed approach significantly reduces communication overhead during training while achieving comparable or even superior performance to that of state-of-the-art supervised methods. 

{\em Notations}: Throughout this paper, upper-case letters (e.g. $X,Y$) represent random variables and lower-case letters (e.g. $\x,\y$) represent the realizations of the corresponding random variables. The shannon entropy of random variable $X$ is denoted as $H(X)$. $I(X;Y)$ is the mutual information between random variables $X$ and $Y$ while $I(X;Y|Z)$ represents the conditional mutual inforamtion between $X$ and $Y$ given $Z$. $D_{KL}(p(\x)||q(\x))$ denotes the Kullback-Leibler (KL) divergence between two probability distributions $p(\x)$ and $q(\x)$.

\section{System Model and Problem Formulation}
\label{sec:system}

\begin{figure*}[t]
    \centering
    \includegraphics[width=1.0\linewidth]{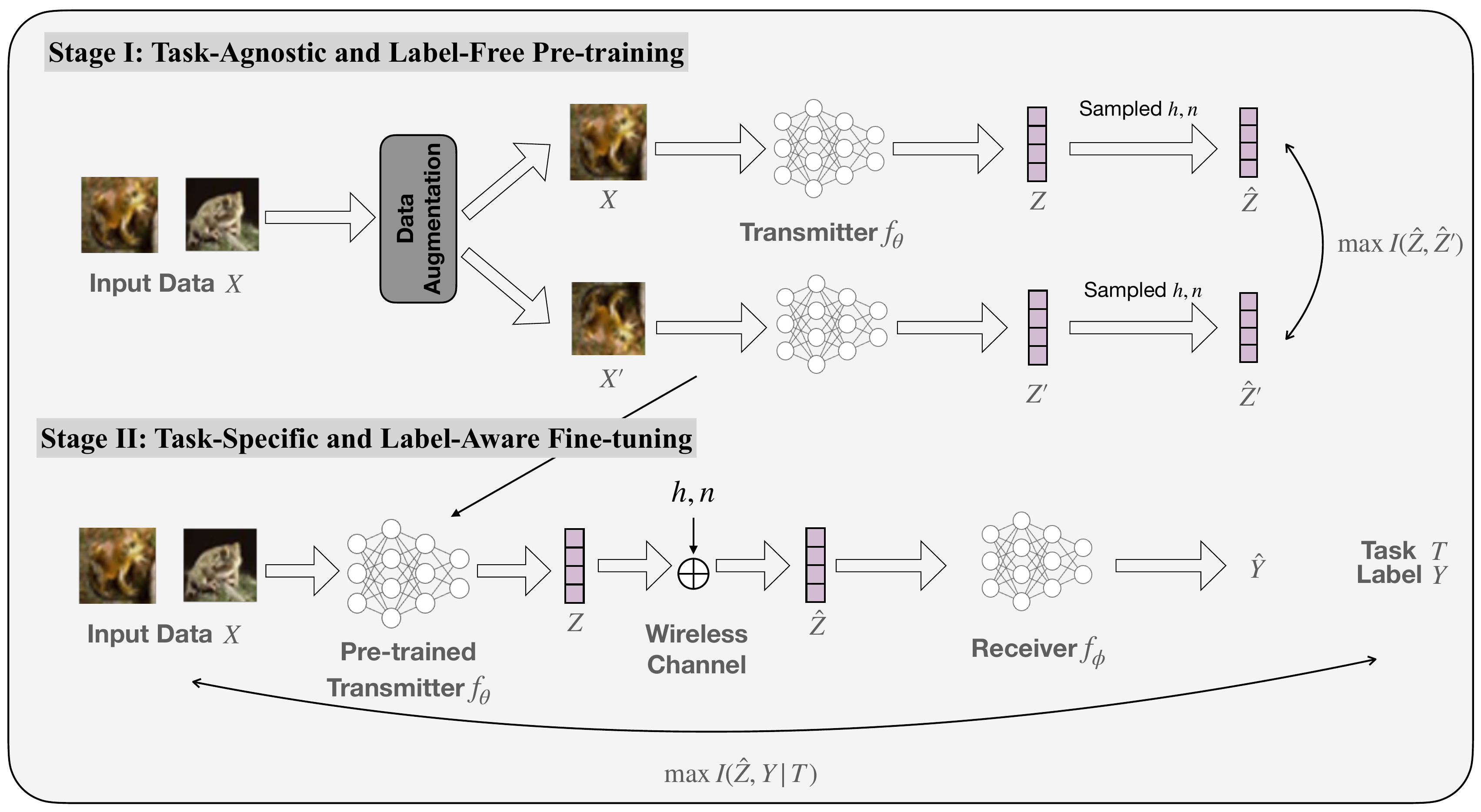}
        \caption{Illustration of our proposed framework for communication-efficient remote training in task-oriented communication. In stage I, the main goal is to learn the parameters $\theta$ at the edge device for feature extraction without the need of the label information $Y$ and the downstream task $T$. Specifically, the edge devices use data augmentation techniques to generate pair data samples $(X,X')$ from the original observed data $X$ and learn task-relevant information from the paired data by maximizing $I(\hat{Z},\hat{Z'})$. Furthermore, we assume the channel statistics are known at the edge devices for local training and the model for the representation extraction can be siamese or dual networks. In stage II, once the inference task $T$ and label $Y$ are known, the edge deivces and the central server jointly train $(\theta,\phi)$ for the inference task by maximizing the representation sufficiency $I(\hat{Z},Y|T)$.}
    \label{fig:sys}
    \end{figure*}

As illustrated in Fig.~\ref{fig:sys}, we consider a task-oriented communication system for edge-device co-inference. 
% The edge device is equipped with sensors thus can observe the environment, while the on-device communication and computation resources are quite limited. 
The edge device can observe the environment and extract features from the raw data, while the on-device communication and computation resources are limited.
The server is equipped with powerful computation resources and makes final inference decision based on the received features. 
% The communication cost of edge device is much higher than the server as the on-device energy consumption is a critical issue in the edge computing system.
% Furthermore, we assume the edge server is not predetermined, as connections between devices and servers are dynamic, with each server potentially requiring different task-relevant features. Consequently, edge devices lack access to label information and downstream task before the connection is established.
Furthermore, we assume the edge server is not predetermined, as connections between devices and servers are dynamic, with each server potentially performing different tasks. Consequently, edge devices lack access to label information and downstream task before the connection is established.
% The probabilistic graphical model of the considered task-oriented communication is 
The probabilistic graphical model (PGM) for the considered task-oriented communication system is given by
\begin{equation}\label{eq:original_graph}
\begin{tikzcd}[row sep=tiny]
    Y&X\arrow[l]\arrow[r]&Z\arrow[r]&\hat{Z}\arrow[r]&\hat{Y},
    \end{tikzcd}
\end{equation}
where the target variable $\boldsymbol{y}$ (i.e., the ground truth for the tasks of interest) and the observations of the edge devices $\x$ are the realization of the random variables $Y$ and $X$ from the joint distribution $p_\mathcal{D}(\boldsymbol{y}, \boldsymbol{x})$. Furthermore, $\mathcal{D}$ denotes the dataset of interest. 
The edge device extracts features $\z$ from the observations $\x$ by the on-device feature encoders $p_{\theta}(\z|\x)$ with the parameter $\theta$. The feature encoders could be deterministic or stochastic. The transmitted signal $\hat{\z}$ is normalized to satisfy $\frac{1}{k}\|\hat{\z}\|^2= P$ where $k$ is the demension of the feature ${\z}$ and $P$ is the power constraint. We use $p_{\mathcal{H}}(\hz|\z)$ to denote the transfer probability of the wireless channel and assume the additive white Gaussian noise (AWGN). The distribution of received feature $\hz$ at the server is represented as
\begin{equation}
    p(\hat{\z}) = p({\z})p_{\mathcal{H}}(\hat{\z}|{\z}),
\end{equation}
where $p(\z)$ is the distribution of the transmitted features $\z$. Specifically, the received feature at the central server is given by
\begin{equation}
    \hat{\z} = \h \odot \z + \n,
\end{equation}
where $\h$ is the channel vector between the edge device and the central server. The opreator $\odot$ denotes the element-wise product and $\n\sim \mathcal{CN}(\boldsymbol{0},\sigma_n^2\boldsymbol{I})$ is the AWGN with variance $\sigma_n^2$. The corresponding signal-to-noise ratio (SNR) is defined as $\text{SNR} = \frac{P}{\sigma_n^2}$. Given the transmitted features $\z$, the conditional probability of $p_{\mathcal{H}}(\hz|\z)$ can be expressed by
\begin{equation}
    p_{\mathcal{H}}(\hat{\z}|{\z}) = \mathcal{CN}(\hat{\z};\h\odot{\z},\sigma_n^2\boldsymbol{I}).
\end{equation}
Then the central server receives the feature $\hat{\z}$ and performs inference based on the received features $\hat{\z}$ by the receiver $p_{\phi}(\hat{\boldsymbol{y}}|\hat{\boldsymbol{z}})$ with parameter $\phi$.

The main design objective is to reduce the overall communication overhead in training stage, given the constraint that the edge devices do not have access to the label information $Y$ and the downstream task $T$ before the connection is established.

\section{Mutual Information Maximization for Task-Agnostic Feature Extraction}
\label{sec:method}
In this section, we develop a two-stage algorithm for accelerating remote training in task-oriented communication, based on a mutual infomration maximization approach. The proposed method consists of two stages: 1) task-agnostic pre-training for transmitter and 2) task-specific fine-tuning for transmitter and receiver. We first introduce the self-supervised feature extraction by data augmentation and utilize a tractable mutual information to maximize the shared information between the data and its augmentation. We then present the overall algorithm for accelerating remote training in task-oriented communication.

\subsection{Self-Supervised Feature Extraction}
\label{sec:method:multi-view}

The minimality and sufficiency of the extracted features are two crucial properties for task-oriented communication systems.
Minimality ensures the features $\hz$ to be as compact as possible to reduce the communication overhead, i.e., $I(X;\hZ)$ should be minimized~\cite{vfe, li2024tackling}. Additionally, the sufficiency requires $\hat{Z}$ is a sufficient representation of $X$ for predicting $Y$ if and only if $I(\hat{Z};Y)=I(X;Y)$.

Among all sufficient representations, the minimal representation is the one that has minimal complexity, i.e., $I(X;Z)$ is minimized. As discussed in~\cite{shao2023task, alemi2016deepVIB,e22090999_ceb}, an ideal representation $\hat{Z}$ should satisfy
\begin{equation}
    \begin{aligned}
        \label{eq:ideal}
        \rlap{$\underbrace{\phantom{{I(X ; \hat{Z}) =I(Y;X)}}}_{\textbf{Minimality}}$} I(X ; \hat{Z}) = \overbrace{I(Y;X) =I(Y ; \hat{Z})}^{\textbf{Sufficiency}},
    \end{aligned}
\end{equation}
and can be achieved by minimizing the IB objective function~\cite{vfe, li2024tackling}.

% However, the IB objective function~\eqref{eq:ideal} requires label information $Y$ for optimizaiton. To address this challenge, we propose a self-supervised feature extraction approach that leverages the multi-view learning framework. 

However, optimizing the IB objective function in \eqref{eq:ideal} requires the known label information $Y$, which is not practical in task-agnostic feature extraction scenarios. To this end, we propose a self-supervised feature extraction approach that leverages a multi-view learning framework~\cite{chen2020simple,tsai2020self}. Inspired by the success of multi-view feature coding for task-oriented communication~\cite{shao2022task}, we reformulate the problem of task-agnostic feature extraction as a multi-view feature extraction problem. 
Specifically, we consider the input data $X$ and its augmented version $X' = Aug(X)$ as two views of the same data. Thus the self-supervised feature extraction problem can be reformulated from the multi-view perspective, e.g., from view $X$ and view $X'$~\cite{chen2020simple,federici2020learning}. We assume the task-relevant information is included in the shared information between the two views, which is reasonable since $X$ and the augmented data $X'$ convey the same semantic content from human perception. Thus the unique information in each view can be considered as the redundant information~\cite{federici2020learning}, allowing us to decompose the mutual information between $X$ and its representation $Z$ as follows,
\vspace{-0.1cm}
\begin{equation}
        \label{eq:redundancy}
        \begin{aligned}
        I(X ; \hat{Z})&\overset{(a)}{=}\ \ \ \ I(X,X';\hat{Z})\\
        &\overset{(b)}{=}\underbrace{I(X ; \hat{Z}| X')}_{\text {\textbf{Redundant Information }}}+\underbrace{I(X' ; \hat{Z})}_{\text {\textbf{Shared Information }}},
        \end{aligned}
    \end{equation}
where (a) holds because $\hat{Z}$ is a noise-corrupted representation of $X$ and makes $\hat{Z}$ conditionally independent of $X'$ given $X$; (b) follows from the chain rule of mutual information.
The redundant information represents the surplus content in the representation $\hat{Z}$ of view $X$ that is not shared by view $X'$, where the shared information is task-relevant that should be extracted. Therefore, to learn the cross-view task-relevant information, we aim to maximize the predictive information $I(X' ; \hat{Z})$. However, directly optimizing $I(X';\hat{Z})$ or find a tractable variational lower bound like that in the IB objective function~\eqref{eq:ideal} is challenging. To address this problem, we propose to maximize the mutual information between the two representations $\hat{Z},\hat{Z'}$ from the two views $X$ and $X'$, i.e., $I(\hat{Z};\hat{Z'})$ as the followling inequality shows,
\begin{equation}
    \begin{aligned}
        I(\hat{Z};\hat{X'})&\overset{(a)}{=}I(\hat{Z};X',\hat{Z'}) - I(\hat{Z};\hat{Z'}|X')\\
        &= I(\hat{Z};X',\hat{Z'})- (H(\hat{Z'}|X')-H(\hat{Z'}|\hat{Z},X'))\\
        &\overset{(b)}{=} I(\hat{Z};X',\hat{Z'})\\
        &\overset{(c)}{=} I(\hat{Z};\hat{Z'}) + I(Z;X'|Z')\\
        &\geq I(\hat{Z},\hat{Z'}),
    \end{aligned}
\end{equation}
where (a) and (c) are obtained by the chain rule of mutual information, and (b) follows from the fact that $\hat{Z'}$ is the received feature of $X'$, thus $\hat{Z'}$ is conditionally independent of $\hat{Z}$ given $X'$. This derivation shows that $I(\hat{Z},\hat{Z'})$ can be considered as a surrogate objective for maximizing the predictive information $I(X';\hat{Z})$. 
To further support the feasiblity of the proposed self-supervised learning framework, similar to~\cite{tsai2020self}, we present the following theorem to analyze the information loss due to wireless channel and internal noise within the dataset.

% We further support the feasiblity of the proposed self-supervised learning framework, we provide the following theorem by analysis the information loss due to wireless channel and internal noise of dataset.

% \begin{theorem}
%     For a specific view $X$ with perfect information transmission, the optimal learned representations from supervised learning satisfy
%     \begin{equation}
%         \hat{Z}^{opt}_{sup}\triangleq \mathop{\arg\max}_{\hat{Z}} I(\hat{Z};Y),
%     \end{equation}
%     Then, given $X$ and $X'=Aug(X)$ with information loss $I(Z;X')-I(\hat{Z};X')=\epsilon_c$ due to the wireless channel, the optimal learned representations from self-supervised learning satisfy
%     \begin{equation}
%         \hat{Z}^{opt}_{ssl}\triangleq \mathop{\arg\max}_{\hat{Z}}I(\hat{Z};X').
%     \end{equation}
%     With the minimal optimal learned representations $\hat{Z}^{opt}_{sup_{min}}$ and $\hat{Z}^{opt}_{ssl_{min}}$ satisfies $\hat{Z}^{opt}_{min}\triangleq \mathop{\arg\min}_{\hat{Z}}H(\hat{Z}|Y)$ for both supervised and self-supervised learning, we have
%     \begin{equation}
%         \begin{aligned}
%             I(X;Y)&=I(\hat{Z}^{opt}_{sup};Y)=I(\hat{Z}^{opt}_{{sup}_{min}};Y)\geq I(\hat{Z}^{opt}_{ssl};Y)\\
%             &\geq I(\hat{Z}^{opt}_{ssl_{min}};Y)\geq I(X;Y)-I(X;Y|X')-\epsilon_c.
%         \end{aligned}
%     \end{equation}
% \end{theorem}

\begin{theorem} 
    For a specific view \( X \) with perfect information transmission, i.e., \( I(Z;X') - I(\hat{Z};X') = 0 \), the optimal learned representations from supervised learning satisfy 
    \begin{equation} 
    \hat{Z}^{\textnormal{opt}}_{\textnormal{sup}} \triangleq \mathop{\arg\max}_{\hat{Z}} I(\hat{Z};Y). 
    \end{equation} 
    Then, given \( X \) and \( X' = \textnormal{Aug}(X) \) with information loss \( I(Z;X') - I(\hat{Z};X') = \epsilon_c \) due to the wireless channel, the optimal learned representations from self-supervised learning satisfy 
    \begin{equation} 
    \hat{Z}^{\textnormal{opt}}_{\textnormal{ssl}} \triangleq \mathop{\arg\max}_{\hat{Z}} I(\hat{Z};X'). 
    \end{equation} 
    With the minimal optimal learned representations \( \hat{Z}^{\textnormal{opt}}_{\textnormal{sup}_{\textnormal{min}}} \) and \( \hat{Z}^{\textnormal{opt}}_{\textnormal{ssl}_{\textnormal{min}}} \) satisfying \( \hat{Z}^{\textnormal{opt}}_{\textnormal{min}} \triangleq \mathop{\arg\min}_{\hat{Z}} H(\hat{Z}|Y) \) for both supervised and self-supervised learning, we have 
    \begin{equation} 
    \begin{aligned} 
    I(X;Y) &= I(\hat{Z}^{\textnormal{opt}}_{\textnormal{sup}};Y) = I(\hat{Z}^{\textnormal{opt}}_{\textnormal{sup}_{\textnormal{min}}};Y) \geq I(\hat{Z}^{\textnormal{opt}}_{\textnormal{ssl}};Y) \\ 
    &\geq I(\hat{Z}^{\textnormal{opt}}_{\textnormal{ssl}_{\textnormal{min}}};Y) \geq I(X;Y) - I(X;Y|X') - \epsilon_c. 
    \end{aligned} 
    \end{equation} 
\end{theorem}
    
\textbf{Theorem 1} shows that the optimal learned representations from self-supervised learning can achieve the same performance as supervised learning with a small information loss. This result further supports the feasibility of the proposed self-supervised learning framework for task-agnostic feature extraction.

\subsection{Tractable Mutual Information Maximization}
The surrogate objective $I(\hat{Z},\hat{Z'})$ is symmetric but challenging to optimize due to the difficulty in estimating high-dimensional mutual information. To address this challenge, we use the InfoNCE~\cite{oord2018representation} to approximate the mutual information between the two representations $\hat{Z}$ and $\hat{Z'}$. Specifically, we maximize the following objective function,
\begin{equation}
    \label{eq:infonce_objective}
    \mathcal{L}_{InfoNCE}(\theta) = \mathbb{E}_{p(x,x')}[\log\frac{e^{<f_\theta(x),f_\theta(x')>}}{\sum_{x_j\in \mathcal{X}}e^{<f_\theta(x),f_\theta(x_j)>}}],
\end{equation}
where $f_\theta(\cdot)$ is the feature extractor parameterized by $\theta$, and $<\cdot,\cdot>$ denotes cosine similarity. 
% The underlying intuition of the InfoNCE estimator is that the cosine similarity between the two representations $\hat{Z}$ and $\hat{Z'}$ from the same data (i.e., the positive pair $(f_\theta(x),f_\theta(x'))$) should be maximized and the cosine similarity between the two representations from different data (i.e., the negative pair $(f_\theta(x),f_\theta(x_j)), for x_j \in \mathcal{X}$) should be minimized to ensure the shared information between the two views can be captured.
The underlying intuition of the InfoNCE estimator is to maximize the cosine similarity between representations $\hat{Z}$ and $\hat{Z'}$ from the same data (i.e., the positive pair $(f_\theta(x),f_\theta(x'))$) while minimizing the cosine similarity between representations from different data (i.e., the negative pair $(f_\theta(x),f_\theta(x_j)), \text{for } x_j \in \mathcal{X}$). This ensures that the shared information between the two views is captured. The InfoNCE estimator provides a lower bound on the mutual information between the representations $Z$ and $\hat{Z}$.
To make this objective tractable, we apply Monte Carlo sampling and temperature scaling, yielding the following practical optimization objective,
\begin{equation}
    \label{eq:infonce_objective_mc}
    \mathcal{L}_{InfoNCE}(\theta)\simeq\frac{1}{2 B} \sum_{i=1}^{2 B}\log \left(\frac{e^{\left(<f_\theta(x_i),f_\theta(x_i')>/\tau\right)}}{\sum_{j=1}^{2B}e^{\left(<f_\theta(x_i),f_\theta(x_j)>/\tau\right)}}\right)
\end{equation}
% where $B$ is the batch size of a mini-batch of data sampled from $\mathcal{X},\mathcal{X'}$, and $\tau$ is the temperature scaling factor. The above objective function can be optimized by stochastic gradient descent (SGD) algorithm thus we can maximize the mutual information between the two views $X$ and $X'$ tractably.
where $B$ is the mini-batch size, and $\tau$ is the temperature scaling factor. This objective function can be optimized using the stochastic gradient descent (SGD) algorithm, enabling a tractable mutual information maximization between the two views $X$ and $X'$.
\subsection{Overall Algorithm}
In previous sections, we have analyzed the feasiblity of learning shared information between two views $X$ and $X'$ for task-agnostic feature extraction and introduced a self-supervised feature approach based on mutual information maximization. Here, we present the overall algorithm designed to accelerate remote training in task-oriented communication. 

The overall algorithm comprises two main stages: 1) task-agnostic pre-training and 2) task-specific fine-tuning. In the pre-training stage, we first pre-train the feature extractor $f_\theta(\cdot)$ with InfoNCE estimator in~\eqref{eq:infonce_objective_mc}, creating a universal feature extractor that captures the intrinsic structure of the input data without requiring downstream task information. In the fine-tuning stage, we update the pre-trained feature extractor in device $f_{\theta}(\cdot)$ and the random inferencer $g_{\phi}(\cdot)$ in server jointly with the task-specific information. The overall algorithm is summarized in Algorithm~\ref{alg:overall}. 

\vspace{5pt}
\begin{algorithm}
    \caption{Self-Supervised Learning for Task-Oriented Communication}
    \label{alg:overall}
    \begin{algorithmic}[1]
        \State \textbf{Input:} Training data $\mathcal{D}=\{x_i,x'_i, y_i\}_{i=1}^{N}$, batch size $B$, temperature scaling factor $\tau$, number of epochs $E_1,E_2$.
        % \State \textbf{Output:} Pre-trained feature extractor $f_\theta(\cdot)$.
        \State \textbf{Pre-training:}
        \State \texttt{\# No information transmission during pre-training}
        \State Initialize $f_\theta(\cdot)$ with random weights.
        \For{$e=1$ to $E_1$}
        \State Sample a mini-batch of data $\{(x_i,x_i')\}_{i=1}^{B}$ from $\mathcal{D}$.
        \State Compute the InfoNCE estimator~\eqref{eq:infonce_objective_mc} with the mini-batch data.
        \State Update the feature extractor $f_\theta(\cdot)$ by minimizing the InfoNCE estimator with SGD.
        \EndFor
        \State \textbf{Fine-tuning:}
        \State \texttt{\# Information transmission is needed during fine-tuning for updating $g_\phi(\cdot)$}
        \State Initialize $f_\theta(\cdot)$ with the pre-trained weights.
        \State Initialize $g_\phi(\cdot)$ with random weights.
        \For{$e=1$ to $E_2$}
        \State Sample a mini-batch of data $\{(x_i,y_i)\}_{i=1}^{B}$ from $\mathcal{D}$.
        \State Compute the task-specific loss with $f_\theta(\cdot)$ and $g_\phi(\cdot)$.
        \State Update $f_\theta(\cdot)$ and $g_\phi(\cdot)$ by minimizing the task-specific loss with SGD.
        \EndFor
    \end{algorithmic}
\end{algorithm}

\section{Performance Evaluation}
\label{sec:simulation}

% We consider the classification task over Ominiglot dataset~\cite{lake2015human} in the following performance evaluation. The Ominiglot dataset contains 1623 different handwritten characters from 50 different alphabets including 964 training characters and 659 testing characters. We consider AWGN channel and Rayleigh fading channel as the wireless channel model. We set the $P=1$ and $\tau =0.1$.
We evaluate performance on a classification task using the Omniglot dataset~\cite{lake2015human}, which contains 1,623 different handwritten characters across 50 alphabets, including 964 training characters and 659 testing characters. For the wireless channel, we consider both the AWGN channel and the Rayleigh fading channel with $P=1$ and $\tau =0.1$.

Performance metrics include test accuracy and task-specific loss (cross-entropy loss for classification) versus the number of communication rounds. One communication round includes both forward and backward passes for a mini-batch data\footnote{As the central server has greater computation and communication resources, we assume no gradient information loss over the wireless channel, with exact gradient recovery at the edge device.}.
\begin{figure*}[t]
    \centering
    \captionsetup[subfigure]{labelformat=empty}
    \subfloat[(a) AWGN, $\textrm{SNR}= 0$ dB]{
        \centering
        \includegraphics[width=0.32\linewidth]{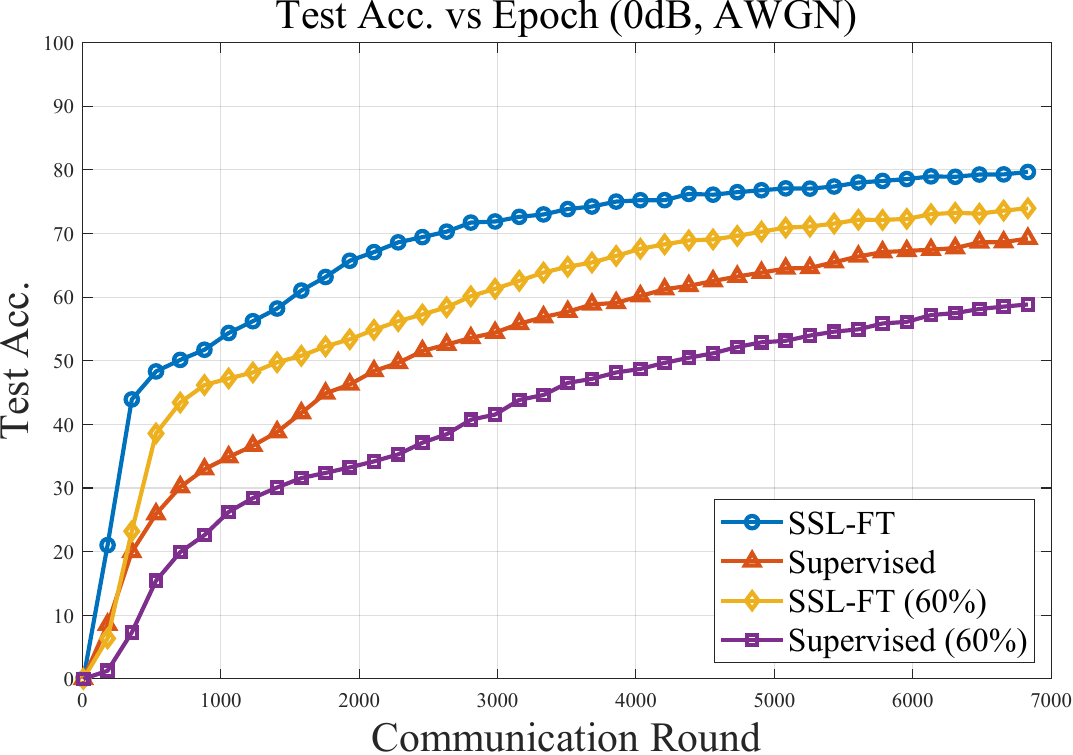}
    }
    \subfloat[(b) AWGN, $\textrm{SNR}= 10$ dB]{
        \centering
        \includegraphics[width=0.32\linewidth]{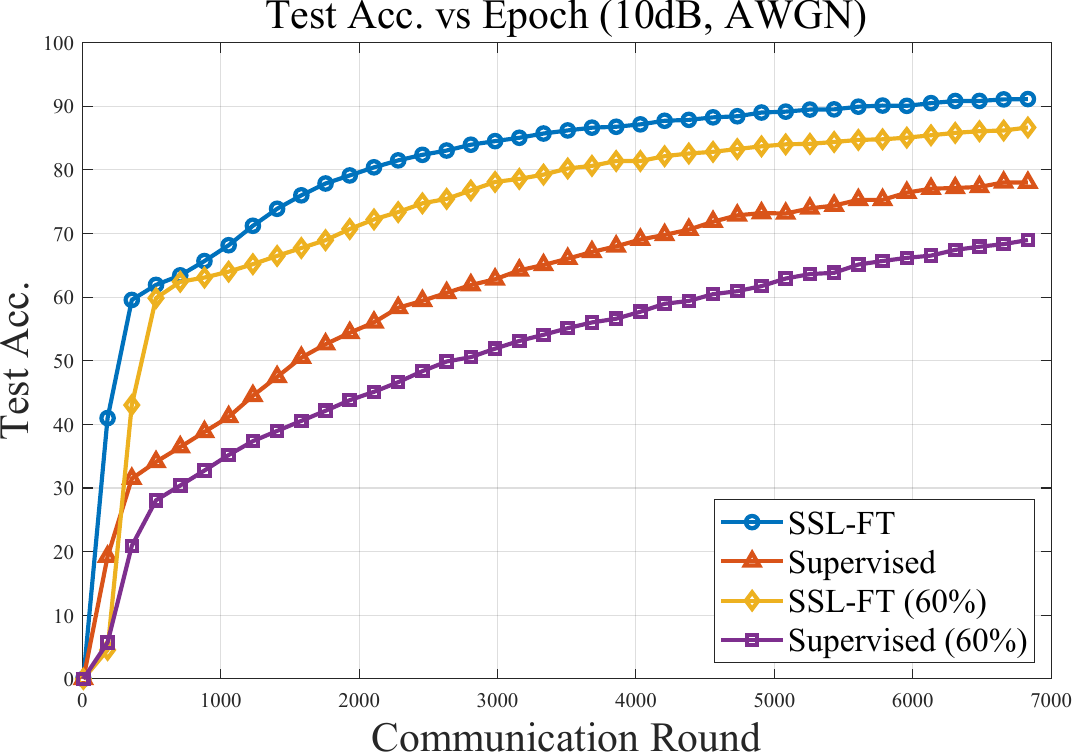}
    }
    \subfloat[(c) AWGN, $\textrm{SNR}= 20$ dB]{
        \centering
        \includegraphics[width=0.32\linewidth]{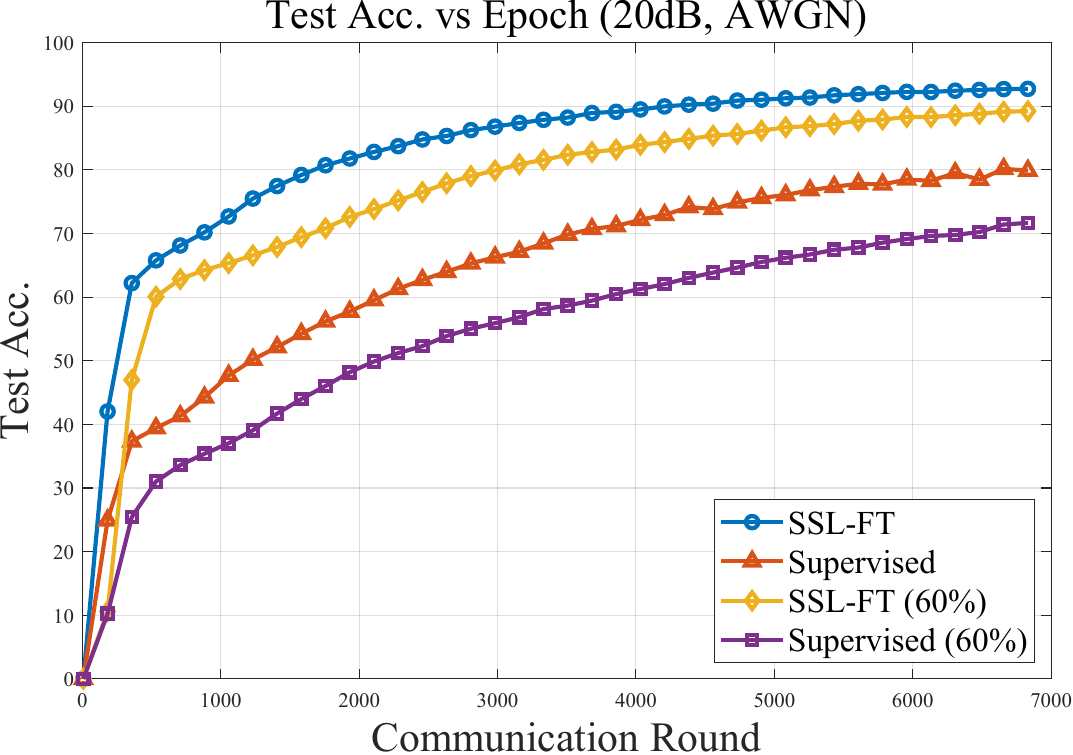}
    }

    \subfloat[(d) Rayleigh, $\textrm{SNR}= 0$ dB]{
        \centering
        \includegraphics[width=0.32\linewidth]{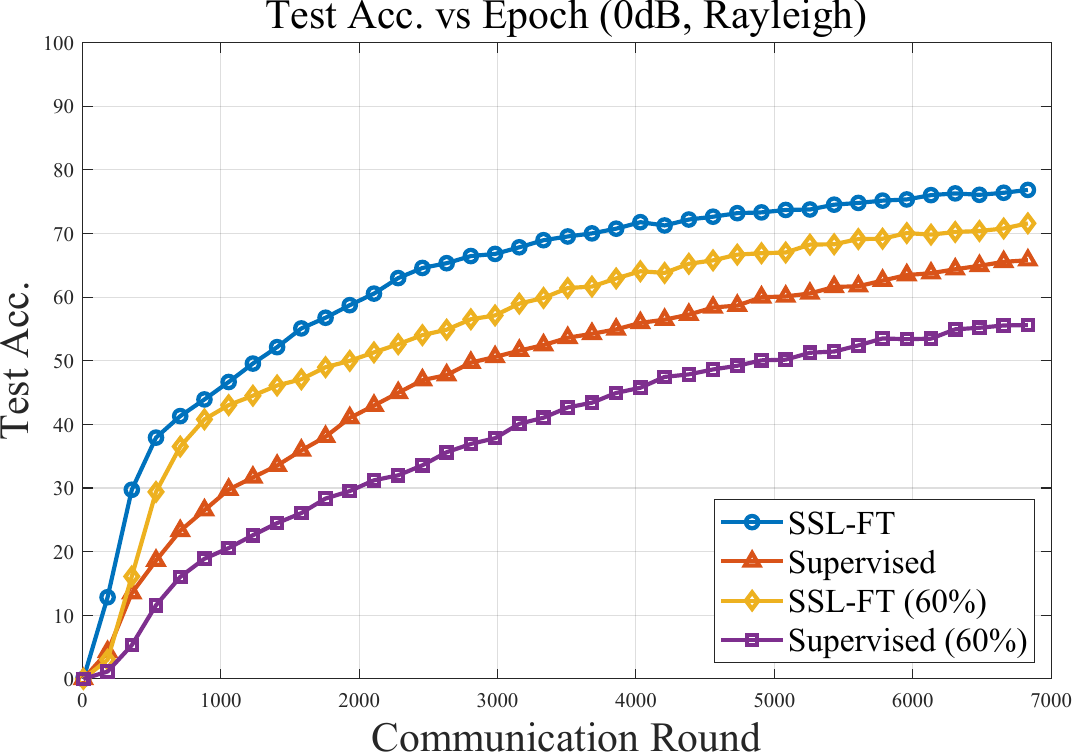}
    }
    \subfloat[(e) Rayleigh, $\textrm{SNR}= 10$ dB]{
        \centering
        \includegraphics[width=0.32\linewidth]{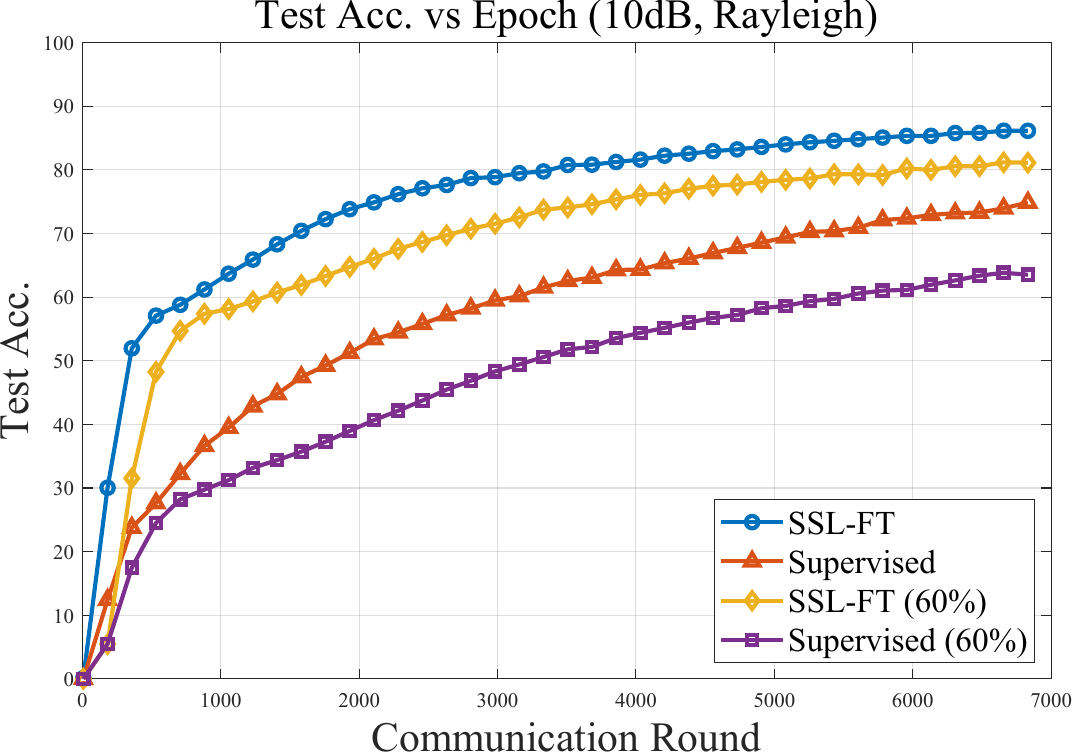}
    }
    \subfloat[(f) Rayleigh, $\textrm{SNR}= 20$ dB]{
        \centering
        \includegraphics[width=0.32\linewidth]{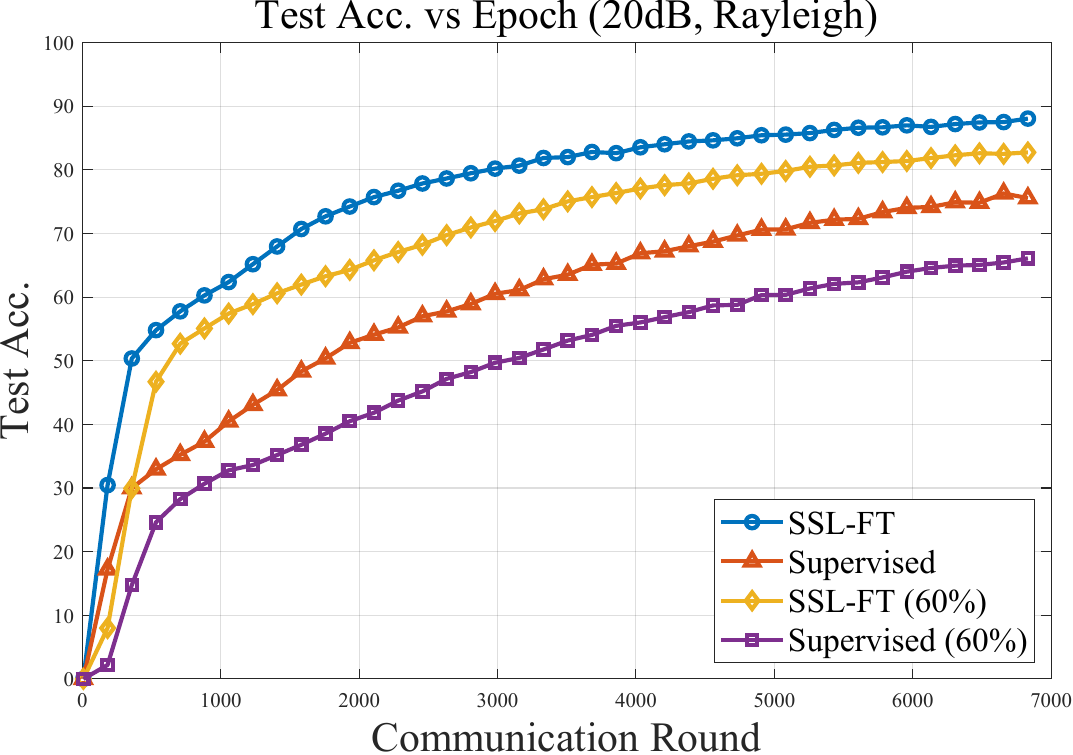}
    }

    % \caption{The rate-distortion curves for image classification tasks with $\text{PSNR}_{\text{test}}=\text{PSNR}_{\text{train}}$. (a) Colored-MNIST, $\textrm{PSNR}= 10\textrm{dB}$, (b) Colored-MNIST, $\textrm{PSNR}= 20\textrm{dB}$, (c) Colored-Object, $\textrm{PSNR}= 10\textrm{dB}$ and (d) Colored-Object, $\textrm{PSNR}= 20\textrm{dB}$.}
    \caption{The communication round versus the test accuracy for the four methods with $\text{SNR}_{\text{test}}=\text{SNR}_{\text{train}}$.}
    \label{fig:results1}
\end{figure*}
We compare the performance of the four methods as follows,
\begin{itemize}
    \item \textbf{Full Supervised (Sup)}: The transmitter and receiver are trained from strach with supervised learning, full label information, and the downstream task.
    \item \textbf{Full Supervised with 60\% Label \stretchrel*{\textbf{(}}{\strut}Sup (60\%)\stretchrel{\textbf{)}}{\strut}}: The transmitter and receiver are trained from strach with supervised learning, 60\% label information, and the downstream task.
    \item \textbf{Self-Supervised and Fine-Tune (SSL-FT)}: The transmitter is pre-trained in a task-agnostic and label-free way, and then fine-tuned with the downstream task and label information.
    \item \textbf{Self-Supervised and Fine-Tune with 60\% Label \stretchrel*{\textbf{(}}{\strut}SSL-FT (60\%)\stretchrel{\textbf{)}}{\strut}}: The transmitter is pre-trained in a task-agnostic and label-free way, and then fine-tuned with 60\% label information and the downstream task.
\end{itemize}
where \textbf{SSL-FT} and \textbf{SSL-FT (60\%)} are our proposed training methods. All methods are trained with the same backbone and optimized using the SGD optimizer with a learning rate of 0.001. The batch size is set to 512, and the total training epochs are 1,000. In each communication round, only one mini-batch is used for training and each epoch-level parameters update requires 7 communication rounds.

In Fig.~\ref{fig:results1}, we illustrate test accuracy versus communication rounds for the four methods under different SNRs. The proposed methods show significantly faster test accuracy growth compared to fully supervised methods, consistently achieving superior test accuracy across all training stages and channel conditions. This is because the proposed methods have captured task-relevant information during the pre-training stage, resulting in faster convergence in the fine-tuning stage. Moreover, with only 60\% of the label information, the proposed \textbf{SSL-FT (60\%)} method outperforms the \textbf{Sup} method that uses full label information, demonstrating the effectiveness of the proposed method in learning with limited labeled data.
% the proposed method's effectiveness in learning with limited labeled data.

Tables I and II shows the results of different methods in terms of loss value under different SNRs. Each element of the table is the minimal communication round needed to achieve a specific training loss; bolded and underlined values indicate the best and second-best performance, respectively. We observe that the proposed methods reach target training loss with fewer communication rounds than fully supervised methods. This is because the proposed methods have already learned a universal representation of input data during the pre-training stage, which can be fine-tuned with fewer labels to perform well on specifc tasks. Additionally, \textbf{SSL-FT (60\%)} even achieves target training loss with fewer communication rounds than the \textbf{Sup} method with full label information, which demonstrates that the proposed method learn more efficiently with less label information.

\begin{table*}[h]
    \centering
    \caption{The communication rounds to achieve different training loss, AWGN Channel}
    \begin{tabular}{@{}c|cccc|cccc|cccc@{}}
    \toprule
    SNR           & \multicolumn{4}{c|}{0dB}   & \multicolumn{4}{c|}{10dB}  & \multicolumn{4}{c}{20dB}  \\ \midrule
    Training Loss & 4.00        & 3.00      & 2.00      & 1.00      & 4.00      & 3.00      & 2.00      & 1.00      & 4.00      & 3.00      & 2.00      & 1.00 \\ \midrule
    SSL-FT        & \bb{560}     & \bb{882}   & \bb{1512}  & \bb{3507}  & \bb{441}   & \bb{679}   & \bb{1099}  & \bb{2100}  & \bb{413}   &\bb{637}    & \bb{1015}  & \bb{1890} \\
    Sup           & \uu{840}     & \uu{1344}  & \uu{2408}  & 6230      & \uu{686}   & \uu{1106}  & 1897      & 4291      & \uu{588}   &\uu{973}    & \uu{1708}  & 3801 \\
    SSL-FT (60\%)  & 980         & 1547      & 2618      & \uu{5957}  & 770       & 1169      & \uu{1855}  & \uu{3514}  & 721       & 1113      & 1750      & \uu{3213} \\
    Sup (60\%)     & 1526        & 2373      & 4144      & 7000      & 1162      & 1890      & 3213      & 6762      & 1029      & 1701      & 2919      & 6055 \\ \bottomrule
    \end{tabular}
\end{table*}

\begin{table*}[h]
    \centering
    \caption{The communication rounds to achieve different training loss, Rayleigh Fading Channel}
    \begin{tabular}{@{}c|cccc|cccc|cccc@{}}
        \toprule
        SNR           & \multicolumn{4}{c|}{0dB}  & \multicolumn{4}{c|}{10dB} & \multicolumn{4}{c}{20dB}  \\ \midrule
        Training Loss & 4.00 & 3.00 & 2.00 & 1.00 & 4.00 & 3.00 & 2.00 & 1.00 & 4.00 & 3.00 & 2.00 & 1.00 \\ \midrule
        SSL-FT        & \bb{658}  & \bb{1022} & \bb{1729} & \bb{4361} & \bb{504}  & \bb{777}  & \bb{1253} & \bb{2590} & \bb{483}  & \bb{749}  & \bb{1218} & \bb{2415} \\
        Sup           & \uu{924}  & \uu{1470} & \uu{2695} & 7000 & \uu{742}  & \uu{1190} & \uu{2086} & 5068 & \uu{672}  & \uu{1120} & \uu{2002} & 4816 \\
        SSL-FT (60\%)  & 1127 & 1750 & 2961 & 7000 & 854  & 1323 & 2121 & \uu{4235} & 868  & 1330 & 2107 & \uu{4095} \\
        Sup (60\%)     & 1589 & 2527 & 4578 & 7000 & 1260 & 2023 & 3472 & 7000 & 1260 & 2016 & 3395 & 7000 \\ \bottomrule
        \end{tabular}
\end{table*}

% \begin{table*}[]
%     \centering

% \end{table*}
\section{Conclusions}
\label{sec:conclusion}
In this paper, we proposed a novel two-stage learning method for task-oriented communication. The proposed method first learns a task-agnostic representation at the transmitter and then fine-tunes the representation with a specific downstream task. 
Simulation results demonstrated that the proposed method can achieve similar performance as existing approaches but with substantially fewer communication rounds, which shows that the proposed method can learn more efficiently with less label information. In the future, it will be essential to extend the proposed method to more complex systems while optimizing both training convergence and overall performance.
% \newpage
% \begin{figure}[t]
%     \centering
%     \includegraphics[width=0.5\linewidth]{image/pgm.pdf}
%     \caption{The probabilistic graphical model of the considered task-oriented communication scheme.}
%     \end{figure}
% \begin{definition}{\textbf{Sufficiency:}}
%     A random variable $Z$ is a sufficient representation of $X$ for predicting $Y$ if and only if $I(Z;Y)=I(X;Y)$.
% \end{definition}

% As discussed in~\cite{shao2023task, alemi2016deepVIB,e22090999_ceb}, an ideal representation $Z_k$ should satisfied
% \begin{equation}
%     \begin{aligned}
%         \rlap{$\underbrace{\phantom{{I(X_{k} ; Z_{k}) =I(Y;X_{k})}}}_{\textbf{Minimality}}$} I(X_{k} ; Z_{k}) = \overbrace{I(Y;X_{k}) =I(Y ; Z_{k})}^{\textbf{Sufficiency}},\ k \in \{1,\ldots,K\},
%     \end{aligned}
% \end{equation}

% \begin{definition}{\textbf{Consistency:}}
    
% \end{definition}

% \begin{definition}{\textbf{Redundancy:}}
%     Given two views $X$ and $X'$, the mutual information between $X$ and its representation $Z$ can be decomposed by
%     \begin{equation}
%         I(X ; Z)=\underbrace{I(X ; Z| X')}_{\text {superfluous information }}+\underbrace{I(X' ; Z)}_{\text {predictive information }},
%     \end{equation}
%     The superfluous part inforamtion is the redundancy between $X$ and $X'$.
% \end{definition}

\appendices
\section{Proof of Theorem 1}

\begin{proof} The proof is mainly based on the chain rule of mutual information and the following PGM, 
    \begin{equation} 
    X' \leftarrow Y \rightarrow X \rightarrow Z \rightarrow \hat{Z}.
    \label{eq:proof_pgm} 
    \end{equation}
    Given label information \( Y \), we have \( \max I(Z^{\text{sup}};Y) = I(X;Y) \). Considering perfect information transmission, we have \( \max I(\hat{Z}^{\text{sup}};Y) = \max I(Z^{\text{sup}};Y) = I(X;Y) \), i.e., \( I(\hat{Z}^{\text{opt}}_{\text{sup}};Y) = I(\hat{Z}^{\text{opt}}_{\text{sup}_{\text{min}}};Y) = I(X;Y) \). With imperfect information transmission, we decompose \( I(\hat{Z}^{\text{ssl}}, X') \) given label information $Y$ by the chain rule of mutual information as follows, \begin{equation} I(\hat{Z}^{\text{ssl}}; X') = I(\hat{Z}^{\text{ssl}}; Y; X') + I(\hat{Z}^{\text{ssl}}; X' | Y). \end{equation} Furthermore, with imperfect information transmission, the information loss is \( I(Z^{\text{ssl}}; X') - I(\hat{Z}^{\text{ssl}}; X') = \epsilon_c \), and according to the PGM in \eqref{eq:proof_pgm}, we have \begin{equation} \max I(\hat{Z}^{\text{ssl}}; X') = \max I(Z^{\text{ssl}}; X') - \epsilon_c = I(X; X') - \epsilon_c. \end{equation} By the chain rule of mutual information, we decompose \( I(\hat{Z}^{\text{ssl}}; X'; Y) \) as \begin{equation} I(\hat{Z}^{\text{ssl}}; X'; Y) = I(\hat{Z}^{\text{ssl}}; X') - I(\hat{Z}^{\text{ssl}}; X' | Y), \end{equation} and have \( I(\hat{Z}^{\text{ssl}}; X' | Y) = 0 \) according to the \( d \)-separation principle. Thus, we have \begin{equation} \max I(\hat{Z}^{\text{ssl}}; X'; Y) = \max I(\hat{Z}^{\text{ssl}}; X') = I(X; X') - \epsilon_c. \end{equation} Then, \begin{equation} \begin{aligned} &I(\hat{Z}^{\text{opt}}_{\text{ssl}}; X') = I(\hat{Z}^{\text{opt}}_{\text{ssl}_{\text{min}}}; X') = I(X; X') - \epsilon_c, \\ &I(\hat{Z}^{\text{opt}}_{\text{ssl}}; X'; Y) = I(\hat{Z}^{\text{opt}}_{\text{ssl}_{\text{min}}}; X'; Y) = I(X; X'; Y) - \epsilon_c. \end{aligned} \label{eq:proof_interMI} \end{equation} By decomposing the second equation in~\eqref{eq:proof_interMI}, we have \begin{equation} \begin{aligned} I(\hat{Z}^{\text{opt}}_{\text{ssl}}; Y) &= I(\hat{Z}^{\text{opt}}_{\text{ssl}}; X'; Y) + I(\hat{Z}^{\text{opt}}_{\text{ssl}}; Y | X') \\ &= -\epsilon_c + I(X; X'; Y) + I(\hat{Z}^{\text{opt}}_{\text{ssl}}; Y | X') \\ &= -\epsilon_c + I(X; Y) - I(X; Y | X') + I(\hat{Z}^{\text{opt}}_{\text{ssl}}; Y | X'). \label{eq:proof_ssl} \end{aligned} \end{equation} A similar derivation can be made for \( I(\hat{Z}^{\text{opt}}_{\text{ssl}_{\text{min}}}; Y) \), \begin{equation} I(\hat{Z}^{\text{opt}}_{\text{ssl}_{\text{min}}}; Y) = -\epsilon_c + I(X; Y) - I(X; Y | X') + I(\hat{Z}^{\text{opt}}_{\text{ssl}_{\text{min}}}; Y | X'). \end{equation} According to the the data processing inequality with \eqref{eq:proof_pgm} and~\eqref{eq:proof_ssl}, we have \begin{equation} \begin{aligned} &I(X; Y | X') \geq I(\hat{Z}^{\text{opt}}_{\text{ssl}}; Y | X'), \\ &I(\hat{Z}^{\text{opt}}_{\text{sup}_{\text{min}}}; Y) = I(\hat{Z}^{\text{opt}}_{\text{sup}}; Y) = I(X; T) > I(\hat{Z}^{\text{opt}}_{\text{ssl}}). \label{eq:proof_comb1} \end{aligned} \end{equation} Furthermore, according to~\eqref{eq:proof_ssl}, \begin{equation} \begin{aligned} I(\hat{Z}^{\text{opt}}_{\text{ssl}_{\text{min}}}) &= -\epsilon_c + I(X; Y) - I(X; Y | X') + I(\hat{Z}^{\text{opt}}_{\text{ssl}_{\text{min}}}; Y | X') \\ &\geq -\epsilon_c + I(X; Y) - I(X; Y | X'). \label{eq:proof_comb2} \end{aligned} \end{equation} Finally, combining~\eqref{eq:proof_comb1} and~\eqref{eq:proof_comb2}, we have \begin{equation} \begin{aligned} I(X; Y) &= I(\hat{Z}^{\text{opt}}_{\text{sup}}; Y) = I(\hat{Z}^{\text{opt}}_{\text{sup}_{\text{min}}}; Y) \geq I(\hat{Z}^{\text{opt}}_{\text{ssl}}; Y) \\ &\geq I(\hat{Z}^{\text{opt}}_{\text{ssl}_{\text{min}}}; Y) \geq I(X; Y) - I(X; Y | X') - \epsilon_c. \end{aligned} \end{equation} \end{proof}

\linespread{0.85}{
\bibliographystyle{./bibtex/IEEEtran}
%\bibliography{/Users/yuanming/OneDrive/Paper/topics/Reference}
% \bibliography{ref}
\bibliography{./bibtex/IEEEabrv,ref}
}
\end{document}